\documentclass[preprint,
superscriptaddress,
groupedaddress,
amsmath,amssymb,
aps,
pre,
floatfix,
]{revtex4-1}
\usepackage{graphicx}
\usepackage{dcolumn}
\usepackage{bm}
\usepackage{wrapfig}
\usepackage{hyperref}

\newcommand{\1}{\'{\i}}

\begin{document}

\title{Response to perturbations as a built-in feature in a mathematical model for paced finger tapping}

\author{Claudia R.\ Gonz\'alez}
\affiliation{Universidad Nacional de Quilmes, Departamento de Ciencia y Tecnolog\1a, Laboratorio de Din\'amica Sensomotora - Bernal, Argentina}
\author{M. Luz Bavassi}
\affiliation{Universidad de Buenos Aires, Instituto de Fisiolog\1a, Biolog\1a Molecular y Neurociencias (IFIByNE), Buenos Aires, Argentina}
\affiliation{CONICET, Buenos Aires, Argentina}
\author{Rodrigo Laje}
\email{Corresponding author: rlaje@unq.edu.ar}
\affiliation{Universidad Nacional de Quilmes, Departamento de Ciencia y Tecnolog\1a, Laboratorio de Din\'amica Sensomotora - Bernal, Argentina}
\affiliation{CONICET, Buenos Aires, Argentina}

\date{\today}

\begin{abstract}

Paced finger tapping is one of the simplest tasks to study sensorimotor synchronization. The subject is instructed to tap in synchrony with a periodic sequence of brief tones, and the time difference (called asynchrony) between each response and the corresponding stimulus is recorded. Despite its simplicity, this task helps to unveil interesting features of the underlying neural system and the error correction mechanism responsible for synchronization. Perturbation experiments are usually performed to probe the subject's response, for example in the form of a ``step change'', i.e.\ an unexpected change in tempo. The asynchrony is the usual observable in such experiments and it is chosen as the main variable in many mathematical models that attempt to describe the phenomenon. In this work we show that although asynchrony can be perfectly described in operational terms, it is not well defined as a model variable when tempo perturbations are considered. We introduce an alternative variable and a mathematical model that intrinsically takes into account the perturbation, and make theoretical predictions about the response to novel perturbations based on the geometrical organization of the trajectories in phase space. Our proposal is relevant to understand interpersonal synchronization and the synchronization to non-periodic stimuli.

\end{abstract}

\pacs{Valid PACS appear here}
%\keywords{Suggested keywords}

\maketitle

\section{Introduction}

Sensorimotor synchronization (SMS), that is the ability to synchronize our movement to a periodic external stimulus, underlies specifically human behaviors like music and dance \cite{Repp2005} and involves time processing in the {\em millisecond timing} range (i.e., hundreds of milliseconds \cite{Ivry2004}). SMS is a rare ability among animals and it apparently correlates with the also rare ability of vocal learning \cite{Schachner2009,Patel2009,Hasegawa2011}, with potential evolutionary implications. The simplest task to study this behavior is paced finger tapping, where a subject is instructed to tap in synchrony with a periodic sequence of brief stimuli (for instance, tones or flashes) like keeping pace with music and while registering the occurrence time of every response. The natural observable and one of the variables most used to quantify the behavior \cite{Chen1997,Repp2013} is the difference between the occurrence times of every response ($R_n$) and the corresponding stimulus ($S_n$), called synchrony error or simply asynchrony:
\begin{equation}
e_n = R_n - S_n \label{eq.observable}
\end{equation}

\noindent The asynchrony of a single trial is a relatively noisy time series with a standard deviation of up to a few tens of milliseconds (Fig.\ \ref{fig.task}). Despite that none of the $e_n$ is zero, most people can achieve average synchronization with a mean value that is typically negative (called Negative Mean Asynchrony or NMA, hypotetically representing the point of subjective synchrony) \cite{Repp2003}. The main goal is to understand how the brain can mantain average synchrony or recover it after a perturbation.

The number of theoretical and experimental works dedicated to understand this behavior and its neural bases is rapidly growing, especially imaging and electrophysiology studies like EEG, MEG, and fMRI \cite{Bavassi2017,Praamstra2003,Pollok2008,Bijsterbosch2011,Nozaradan2018,Jantzen2018,Iversen2016,Merchant2015}. It is very simple to show that there is an error correction mechanism in the brain in charge of keeping average synchrony that operates based on past performance \cite{Bavassi2013,Repp2005}. On the theoretical side \cite{Bavassi2013,vanderSteen2013}, such a mechanism is easily conceptualized as a map or difference equation for the observable $e_n$:
\begin{equation}
e_{n+1} = f(e_n,T_n) + \mbox{noise} \label{eq.oldmodel}
\end{equation}

\noindent where the asynchrony at the next step $e_{n+1}$ depends on its previous value $e_n$ (or several previous values in some models \cite{Pressing1998}) and probably on some parameter like the sequence period or interstimulus interval $T_n$:
\begin{equation}
T_n = S_n - S_{n-1}.
\label{eq.period}
\end{equation}

Studies that aim at finding a linear correction function $f$ make use of mean values, standard deviations and auto-correlation functions and thus they analyze synchronization to periodic stimuli sequences without any perturbation \cite{Pressing1998}; periodic sequences are also used by works that study the structure of the noise term \cite{Wagenmakers2004}. Alternatively, in order to find the best correction function $f$ that is the deterministic part of the equation one can perform perturbations to an otherwise isochronous sequence and analyze the resynchronization \cite{Bavassi2013}. That is the approach we chose.

The main variable for quantifying the behavior, the asynchrony $e_n$, is always operationally well defined according to Eq.\ \ref{eq.observable} and Fig.\ \ref{fig.task}. In this work, however, we show that $e_n$ is an ill-defined variable in terms of a map or difference equation when the stimulus sequence has perturbations. This issue is relevant not only during a controlled perturbation experiment in a laboratory setting, but also in a more natural, ecological setting like music performance where the stimulus sequence is not strictly periodic (e.g., a choir or orchestra conductor performing a ralentando) or when the stimulus sequence is intrinsically variable (e.g., interpersonal synchronization where the stimulus sequence is the other person's production). We propose an alternative variable and a mathematical model that reproduces the observed data including the effect of perturbation and make theoretical predictions.

\section{Results}
\subsection{Predicted versus observed asynchrony} \label{sec.predicted}

\begin{figure}[t]
    \centering
    \includegraphics[width=0.5\linewidth]{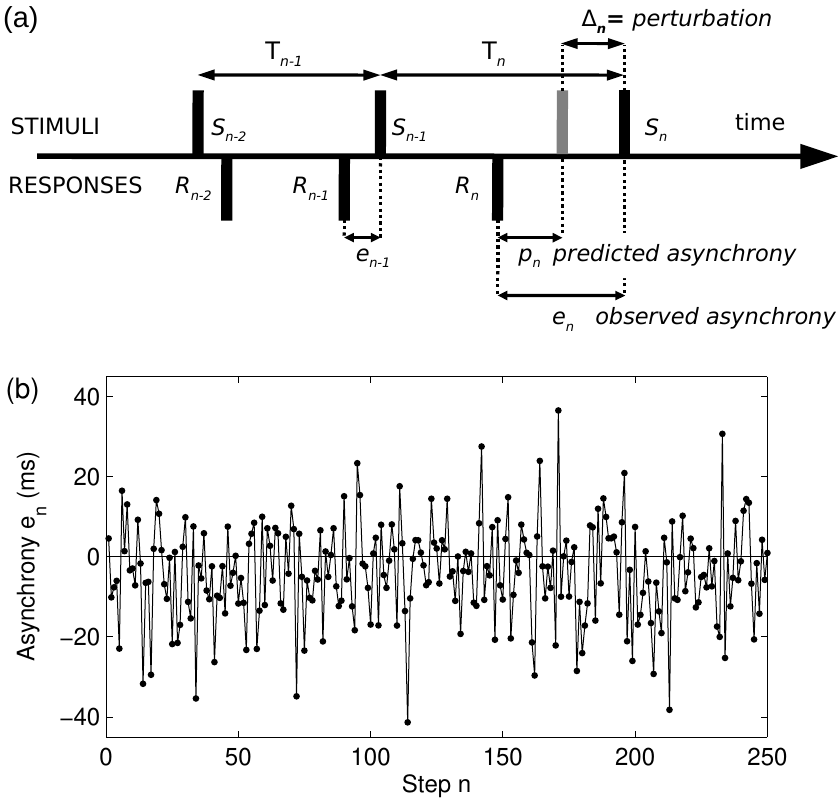}
    \caption{(a) Schematic description of the task, variables and parameters. A perturbation occurs at step $n$. (b) Isochronous sequence from one subject (single trial, no averaging). Mean = $3.9$ ms, SD = $12.1$ ms.}
    \label{fig.task}
\end{figure}

From a behavioral point of view, one of the approaches to unveil the form of the error correction mechanism in a paced finger tapping task is to find the best correction function $f(e_n)$ (Eq.\ \ref{eq.oldmodel}) that, based on the observed asynchrony $e_n$ (Ec.\ \ref{eq.observable}) and the interstimulus interval $T_n$ (Eq.\ \ref{eq.period}), predicts the asynchrony at the next step. Figure \ref{fig.task} shows a graphical definition of all variables and parameters.

According to Eq.\ \ref{eq.observable}, the asynchrony $e_n$ takes as a reference point the occurrence time of the stimulus $S_n$. The traditional way of perturbing the system is to unexpectedly modify the stimulus period, which is done by shifting in time one or several consecutive stimuli, i.e.\ modifying $S_n$ (and perhaps the following stimuli too; see Fig.\ \ref{fig.task}). An example is the ``step change'' perturbation where the stimulus period changes unexpectedly by a fixed amount $\Delta$ from a given stimulus $S_n$ on (in musical terms it is a sudden change in tempo). A critical issue that has been overlooked in the literature, both experimental and theoretical, is that the change of the occurrence time of the perturbed stimulus $S_n$ is arbitrary and thus is not a well defined time reference \cite{Lopez2019}. The consequence of this is that the variable $e_n$ becomes ill-defined because its value changes at the moment of perturbation but not because of its own dynamics. In other words, if an unexpected perturbation occurs at step $n$, the actual asynchrony will be different from the value predicted by the subject (or the model) because the corresponding stimulus was shifted by an arbitrary amount equal to the size of the perturbation.

In this work we propose the following way to resolve this issue: we distinguish between the {\em predicted} asynchrony value $p_n$ and the actually {\em observed} asynchrony value $e_n$ (Fig.\ \ref{fig.task}). If a change in period occurs at step $n$:
\begin{equation}
\Delta_n = T_n - T_{n-1} \label{eq.delta}
\end{equation}

\noindent then the predicted asynchrony $p_n$ and the actually observed asynchrony $e_n$ are related by the following expression:
\begin{equation}
p_n =  e_n + \Delta_n \label{eq.predicted_observed}
\end{equation}

\noindent It is important to note that Eq.\ \ref{eq.predicted_observed} is not a theoretical assumption but the actual relationship between the variables. Note also that if no perturbation occurs then $\Delta_n = 0$ and thus $p_n = e_n$.

Taking this distinction into account, a model like Eq.\ \ref{eq.oldmodel} must be written as (discarding the noise term for simplicity):
\begin{equation}
p_{n+1} = f(e_n,T_n) \label{eq.newmodel}
\end{equation}

\noindent and by using Eq.\ \ref{eq.predicted_observed} we can get a closed model for the variable $p_n$:
\begin{equation}
p_{n+1} = f(p_n - \Delta_n,T_n)
\end{equation}

\noindent We can always go back to the observed asynchrony $e_n$ by means of Eq.\ \ref{eq.predicted_observed}.

\subsection{Model implementation}

\begin{figure}[t]
    \centering 
    \includegraphics[width=0.5\linewidth]{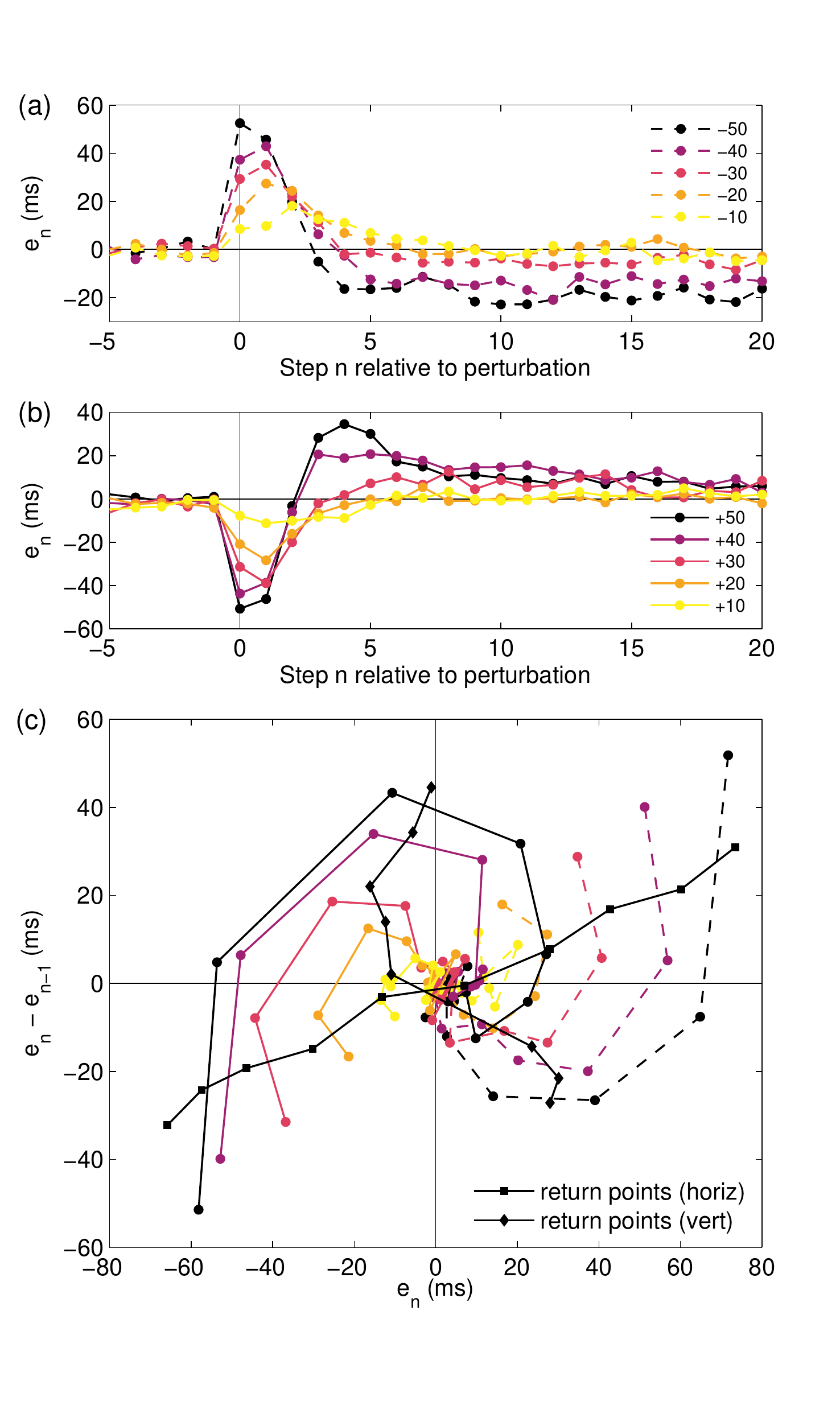}
    \caption{Experimental time series and phase space reconstruction (all experimental data shown in this work were published in \cite{Bavassi2013}). Perturbation sizes are in milliseconds and the pre-perturbation period is $T_0=500$ ms. (a) Negative step changes. (b) Positive step changes. (c) Phase space reconstruction by embedding the time series above. Note that there is a common region of phase space shared by trajectories from large opposite perturbations (most notably $+50$ and $-20$ trajectories). Black curves are the estimated return points.}
    \label{fig.old_expdata}
\end{figure}

In this section we proposed an improved model based on previous work \cite{Bavassi2013} and apply the distinction proposed above. In \cite{Bavassi2013} we showed that nonlinear effects are important even for small perturbations of 10\% of the period. There are two main effects when step change perturbations are performed. If the tempo is made faster (negative perturbation $\Delta_n<0$; Fig.\ \ref{fig.old_expdata}(a)) the resynchronization is monotonic until a new baseline is reached; however, if the tempo is made slower (positive perturbation $\Delta_n>0$; Fig.\ \ref{fig.old_expdata}(b)) the asynchrony overshoots before reaching the new baseline. The existence of the overshoot makes necessary to introduce a second variable $x_n$---otherwise the deterministic nature of the model would be violated \cite{Bavassi2013,Schoner2002}. In addition, the asymmetry of the response in front of symmetric perturbations requires the introduction of nonlinear terms of even order (e.g.\ quadratic). In order to correctly reproduce these two observations we proposed a two-variable, nonlinear model \cite{Bavassi2013}:
\begin{equation}
\begin{aligned}
e_{n+1} & = a \, e_n + b (x_n - T_n) + F(e_n,x_n,T_n) \\
x_{n+1} & = c \, e_n + d (x_n - T_n) + T_n + G(e_n,x_n,T_n)
\end{aligned} \label{eq.Bavassi_model}
\end{equation}

\noindent where $F$ and $G$ are nonlinear functions of their arguments.

Before introducing the distinction between predicted and observed asynchrony, we propose improved functions $F$ and $G$ based on new experimental evidence. We take the experimental data of step change perturbations from \cite{Bavassi2013} and reanalyze them as follows. First we perform an embedding of the time series shown in Figs.\ \ref{fig.old_expdata}(a) and \ref{fig.old_expdata}(b) to reconstruct the qualitative geometrical arrangement of the underlying phase space \cite{Gilmore1998}. Figure\ \ref{fig.old_expdata}(c) shows the result, where we added the detection of return points for every trajectory (see methods in Appendix\ \ref{Estimation of return points}). The geometrical organization of the return points shows a certain degree of asymmetry and saturation (especially for the vertical axis), which tells us that quadratic and cubic terms are needed. There are six possible quadratic terms and eight possible cubic terms, and we want to use the smallest number of terms that correctly represent the behavior. However, we cannot use normal form theory to choose among the nonlinear terms because within the regime analyzed in this work (synchronization to a periodic sequence or resynchronization following a tempo step change of fixed size) the behavior does not show any bifurcations but a robust convergence to a single fixed point representing average synchrony \cite{Bavassi2013,Thaut1998,Repp2001,Repp2004}. After testing many combinations of nonlinear terms with qualitatively similar results, we choose the following:
\begin{equation}
\begin{aligned}
F(e_n,x_n,T_n) & = \alpha e_n^3 + \beta e_n (x_n-T_n)^2 + \gamma (x_n-T_n)^3 \\
G(e_n,x_n,T_n) & = \delta e_n^2
\end{aligned} \label{eq.newmodel_NLterms}
\end{equation}

\noindent In Section \ref{sec.fitting} we display a summary of the obtained phase spaces with our selection of nonlinear terms. We emphasize that our choice of nonlinear terms is not unique---it is not possible to solve for unique $F$ and $G$ from the shape of the return points in the embedding.

Now we incorporate the distinction between predicted and observed asynchrony. According to the previous subsection we must substitute $e_{n+1} \rightarrow p_{n+1}$ and $e_n \rightarrow p_n - \Delta_n$:
\begin{equation}
\begin{aligned}
p_{n+1} & = a \, (p_n-\Delta_n) + b (x_n - T_n) + {} \\
& \hspace{1cm} + F(p_n-\Delta_n,x_n,T_n) \\
x_{n+1} & = c \, (p_n-\Delta_n) + d (x_n - T_n) + T_n + {} \\
& \hspace{1cm} + G(p_n-\Delta_n,x_n,T_n)
\end{aligned} \label{eq.newmodel_a}
\end{equation}

\noindent and to get a closed system we define the auxiliary variable $s_{n+1} = T_n$. This leads us to
\begin{equation}
\begin{aligned}
p_{n+1} & = a \, (p_n-(T_n-s_n)) + b (x_n - T_n) + {} \\
& \hspace{1cm} + F(p_n-(T_n-s_n),x_n,T_n) \\
x_{n+1} & = c \, (p_n-(T_n-s_n)) + d (x_n - T_n) + T_n + {} \\
& \hspace{1cm} + G(p_n-(T_n-s_n),x_n,T_n) \\
s_{n+1} & = T_n
\end{aligned} \label{eq.newmodel_b}
\end{equation}

\noindent where $F$ and $G$ are defined according to Eq.\ \ref{eq.newmodel_NLterms}. It is worth noting that $T_n$ is a parameter whose value is up to the experimenter.

\subsection{Model fitting, model simulations, and fitting analysis}
\label{sec.fitting}

\begin{table}[b]
\centering
\begin{tabular}{rclp{0.25cm}|p{0.25cm}rcl}
$a$ & = & $-0.0485$ & & & $\alpha$ & = & $5.67 \times 10^{-5}$ \\
$b$ & = & $0.467$ & & & $\beta$ & = & $7.71 \times 10^{-5}$ \\
$c$ & = & $-0.491$ & & & $\gamma$ & = & $9.74 \times 10^{-5}$ \\
$d$ & = & $0.987$ & & & $\delta$ & = & $4.61 \times 10^{-3}$
\end{tabular}
\caption{Fitted parameter values. All simulations in this work use this unique set of parameter values unless stated otherwise. Linear coefficients $a$, $b$, $c$, $d$ are nondimensional; quadratic coefficient $\delta$ has units of ms$^{-1}$; cubic coefficients $\alpha$, $\beta$, $\gamma$ have units of ms$^{-2}$. The pre-perturbation interstimulus interval is $T_0 = 500$ ms.}
\label{tab.param_values}
\end{table}

As we did before \cite{Bavassi2013}, we use a genetic algorithm to fit the model (Eq.\ \ref{eq.newmodel_b}) to the experimental time series of $\Delta = \pm50$ ms only (Fig.\ \ref{fig.model_simuls}(a); see methods in Appendix\ \ref{Genetic algorithm}). The obtained parameter values are shown in Table \ref{tab.param_values} and the corresponding phase space is shown in Fig.\ \ref{fig.model_simuls}(b).

The fitting goodness is noteworthy, especially when compared to previous attempts in the literature with a similar or even greater number of parameters (for instance our own previous work \cite{Bavassi2013} and many others' work, e.g.\ \cite{Thaut1998,Repp2001,Schulze2005,Large2002,Repp2004,Repp2012}). It must be taken into account that, contrary to ours, modeling and fitting efforts in the finger tapping literature usually perform separate fits for different perturbation sizes, effectively multiplying the number of parameters by a factor of 2 at least.

\begin{figure}[t]
    \centering
    \includegraphics[width=0.5\linewidth]{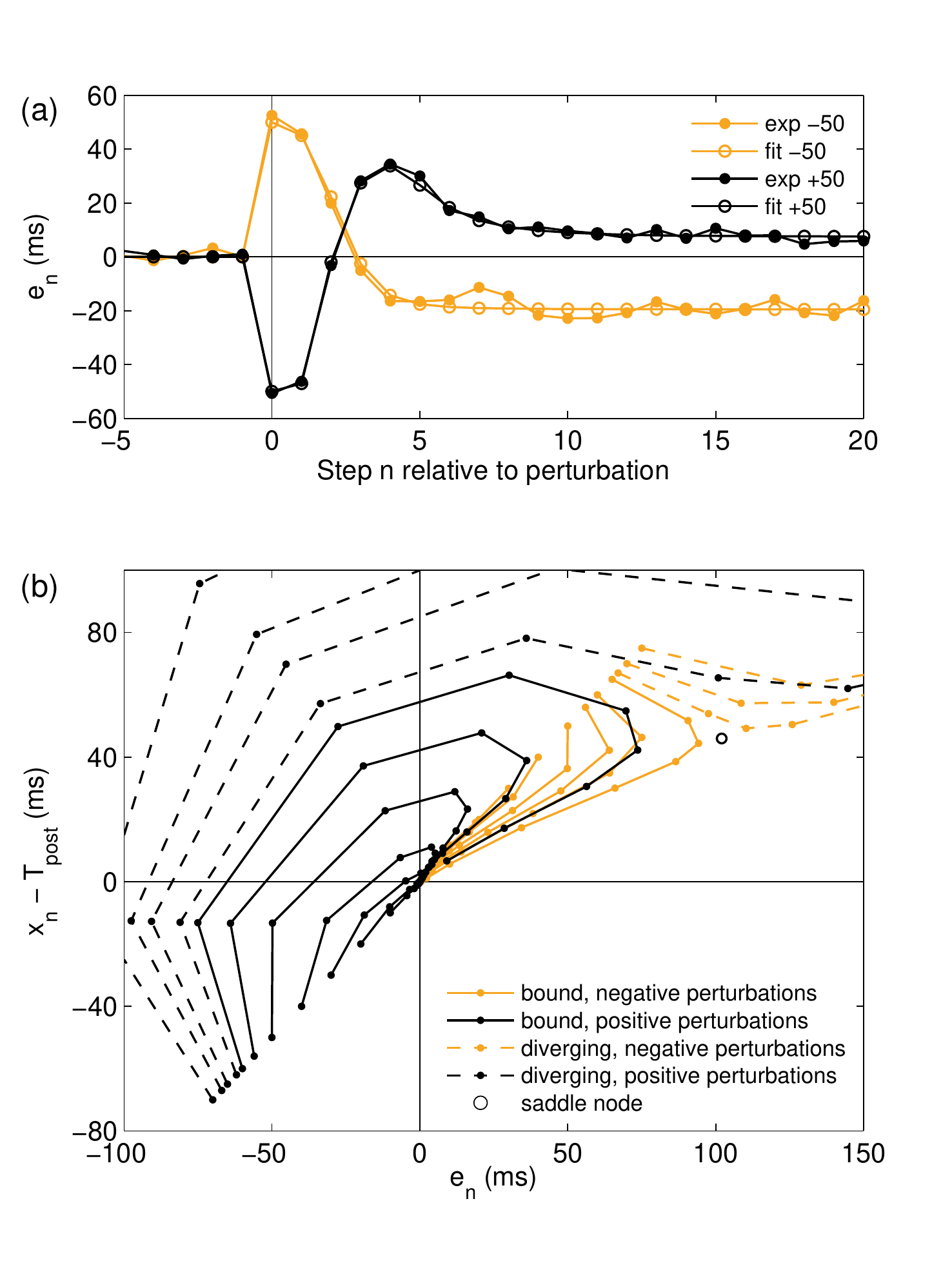}
    \caption{Model fitting. (a) Experimental $\pm 50$ ms time series and model fitting results. (b) Model phase space for tempo step change perturbations. Larger perturbations begin farther away from the origin (preperturbation steps not shown for visual clarity). Note the existence of a saddle node in the upper right quadrant---if the perturbation size is large enough on either side the trajectory diverges (dashed lines), meaning that the subject is not able to resynchronize.}
    \label{fig.model_simuls}
\end{figure}

\begin{figure*}[t]
    \centering
    \includegraphics[width=\textwidth]{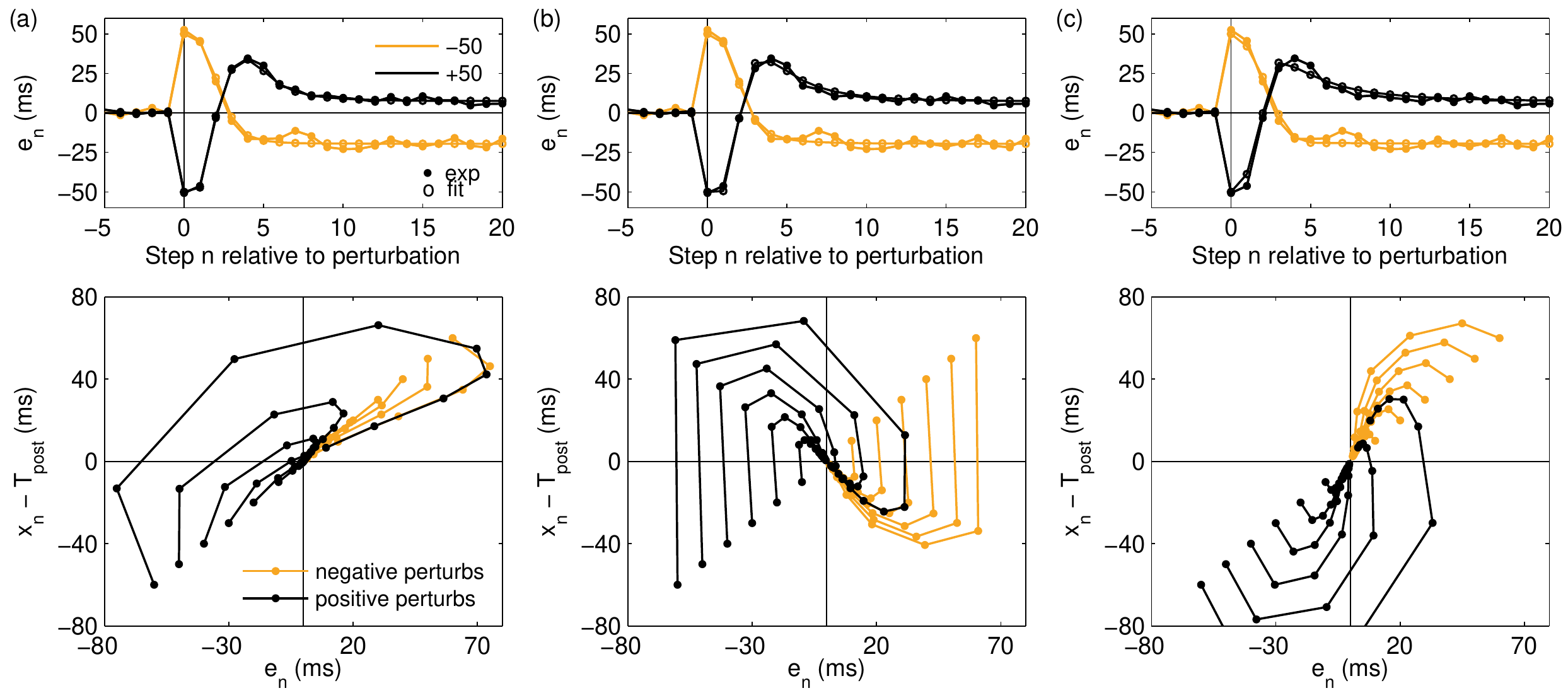}
    \caption{Variety of solutions from the fitting procedure. (a) Type-I phase space and time series (64\% of all solutions; the fittest solution of all, shown in Table \ref{tab.param_values} and Figs.\ \ref{fig.model_simuls}, \ref{fig.benefits}, \ref{fig.model_predictions1}, and \ref{fig.model_predictions2}, is this type. (b) Type-II phase space and time series (25\% of all solutions;  parameter values in Appendix\ \ref{Parameter values of Type-II and Type-III phase spaces}). (c) Type-III phase space and time series (11\% of all solutions; parameter values in Appendix\ \ref{Parameter values of Type-II and Type-III phase spaces}.}
    \label{fig.types}
\end{figure*}

In order to analyze the robustness of the fitting results we proceeded as follows. Every run of a genetic algorithm provides a solution that best fits the data according to the fitness function. As we describe in the Appendix\ \ref{Genetic algorithm} (also in our previous work \cite{Bavassi2013}) we decided to run the algorithm 200 times to choose the absolute best among those 200 converged solutions. This, in addition, allows us to perform a statistical and dynamical analysis of all solutions. The distributions of obtained values for every parameter can be seen in Appendix\ \ref{Fitted parameter distributions}. All distributions are mostly unimodal, which speaks in favor of the robustness of the fitting procedure (with a slight bimodality in some of them). On the other hand, there is some interdependency between some of the linear parameters (i.e., a correlation between their converged values, for instance between $a$ and $d$, but not between nonlinear parameters, but a small correlation among coefficient $\alpha$ and coefficients $a$ and $d$ (see Appendix\ \ref{Fitted parameter distributions}). This means that the model might be overparameterized and might fit the experimental time series with a smaller number of linear terms, but our choice of nonlinear terms is appropriate.

We plotted the phase space of every obtained solution and made an exhaustive visual search in order to qualitatively classify them. We found three types of phase spaces (Fig.\ \ref{fig.types}). The phase space we chose for this work (Figs.\ \ref{fig.model_simuls}(b) and \ref{fig.types}(a)) is the fittest of all 200 solutions and in addition belongs to the most frequent type of phase space (type I); the parameter values corresponding to phase space types II and III are in Appendix\ \ref{Parameter values of Type-II and Type-III phase spaces}).

\subsection{Benefits of the proposed approach}

The most important feature of our model is that it incorporates the perturbations to the stimulus period in the modeling approach, leading to autonomous dynamics once the stimulus period sequence $T_n$ is chosen as input. We achieved this by developing a closed equation for variable $p_n$ (Eq.\ \ref{eq.newmodel_b}) based on a model-free relationship between $p_n$ and the observable $e_n$ (Eq.\ \ref{eq.predicted_observed}). An advantage of having a closed model with built-in perturbations is the ability to perform a bifurcation analysis on it, which is the subject of future work.

\begin{figure*}[t]
    \centering
    \includegraphics[width=\textwidth]{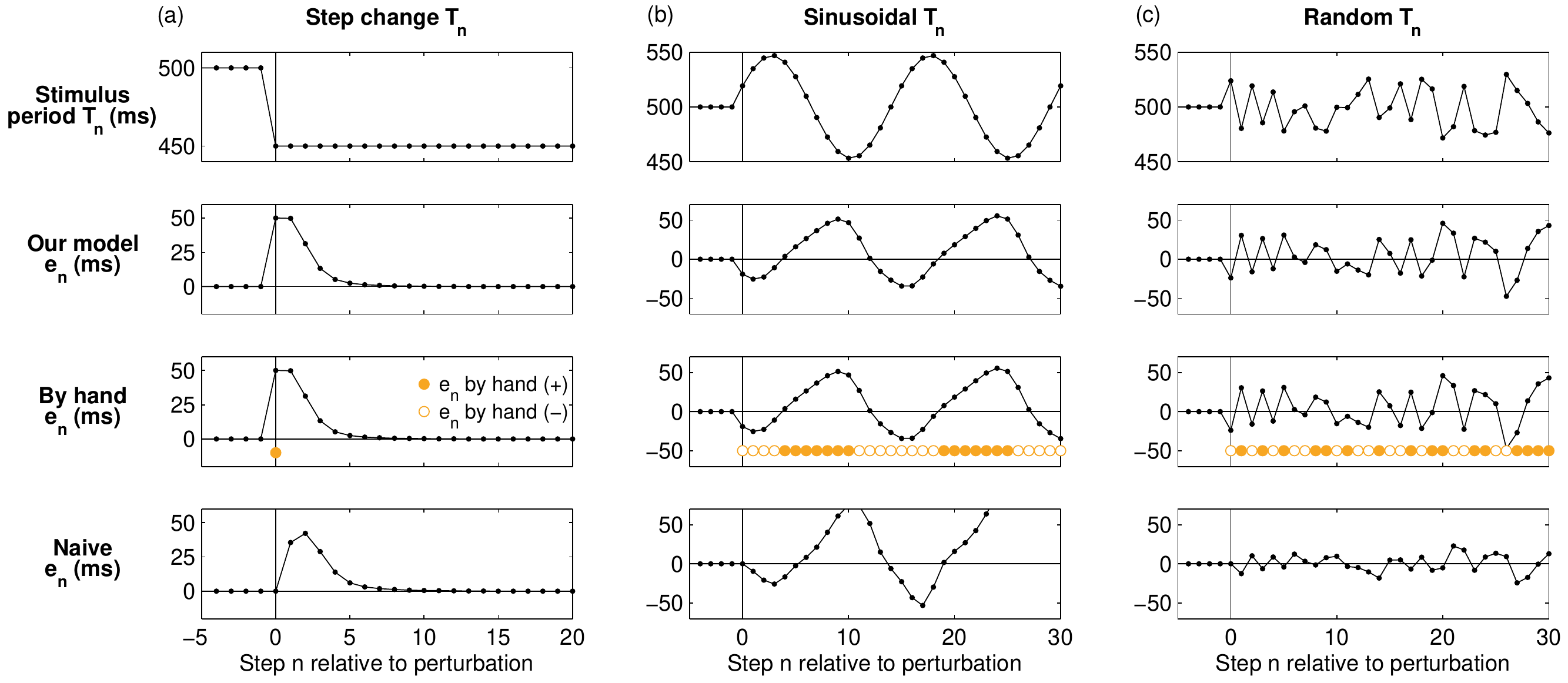}
    \caption{Three common experimental manipulations of the stimulus period (top row) and response of our model (second row). (a) Step change in stimulus period. (b) Sinusoidal change. (c) Pseudorandom change. For comparison, the third row shows the behavior of the model when no distinction between predicted $p_n$ and observed $e_n$ is made, that is the traditional way of modeling where by-hand modification of $e_n$ is needed (shown as orange filled circles when $e_n$ must be increased and open circles when it must be decreased); fourth row shows a ``naive'' approach where neither distinction $p_n/e_n$ nor by-hand modification are made.}
    \label{fig.benefits}
\end{figure*}

In the absence of bifurcations, as is the case in this work, our approach still offers advantages in the form of autonomous dynamics without any need to modify the value of the variable ``by hand'' whenever the stimulus period changes. This is illustrated in Fig.\ \ref{fig.benefits}, where we show three common experimental manipulations of the stimulus period $T_n$ (panel (a), step change; panel (b) sinusoidal variation; panel (c) random variation). Once the stimulus period sequence is set (top row), our model evolves in time without any intervention from the experimenter (second row). The traditional way of doing this (third row) consists in considering the model without distinguishing between $p_n$ and $e_n$, that is Eq.\ \ref{eq.Bavassi_model}, but this of course needs adjusting the value of the variable $e_n \rightarrow e_n - \Delta T$ by hand (shown in the figure as orange circles) every time there is a change in the parameter $T_n \rightarrow T_n + \Delta T$ to produce the correct time evolution (in the sinusoidal and random variations the experimenter needs to adjust $e_n$ at every step). Finally, a naive version is shown in the bottom row where Eq.\ \ref{eq.Bavassi_model} is used without adjusting the value of $e_n$ by hand.

Note that we decided to plot $e_n$ instead of $p_n$ throughout this work mainly for historical reasons so it is easier to compare to previous models' results and experimental results. Keep in mind, however, that all numerical simulations of our proposed model in this work were made by solving the closed equation for $p_n$, Eq.\ \ref{eq.newmodel_b}, and then translated $p_n \rightarrow e_n$ by means of Eq.\ \ref{eq.predicted_observed}.

\subsection{Theoretical predictions}

\subsubsection{Geometrical organization of phase space}

\begin{figure*}[t]
    \centering
    \includegraphics[width=\textwidth]{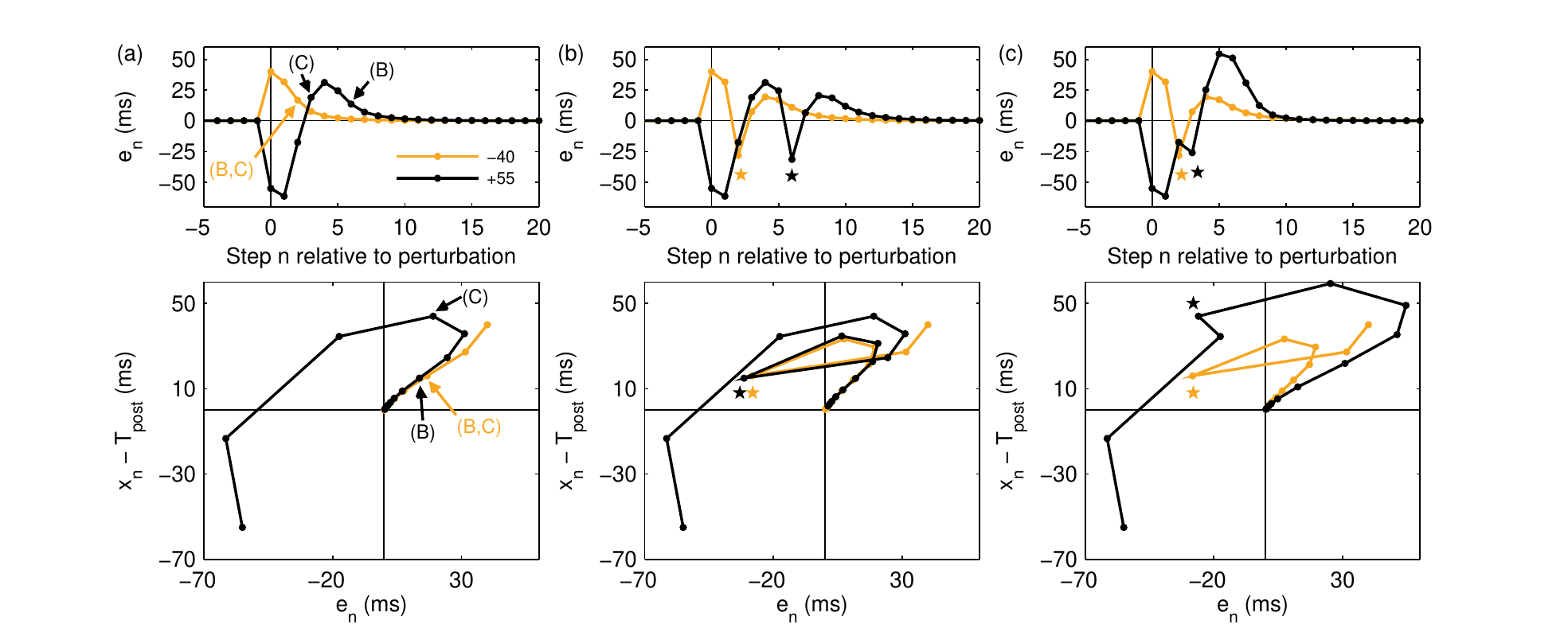}
    \caption{Model predictions (I). (a) Response to a step change perturbation and choice of points for the second perturbation. Points labeled ``B'' and ``C'' have similar asynchrony values and correspond to the perturbations in panels (b) and (c), respectively. (b) Response to the proposed manipulation: both conditions of perturbation 1 give similar responses to perturbation 2. The perturbed points are labeled with stars and correspond to the ``B'' label in panel A. (c) Response to consecutive perturbations as in (b) but the second perturbation takes place in different points of phase space: the response after perturbation 2 depends on the condition of perturbation 1, despite having similar initial asynchrony values (positive control). The perturbed points are labeled with stars and correspond to the ``C'' label in panel (a).}
    \label{fig.model_predictions1}
\end{figure*}

The geometrical arrangement of the trajectories in the experimental phase space is remarkable (Fig.\ \ref{fig.old_expdata}(c)). The trajectories corresponding to the largest positive perturbations share a region of phase space during the overshoot with some of the trajectories of the negative perturbations (that do not have overshoot). This suggests that the error correction mechanism in the brain might not distinguish between both types of trajectories while they are roughly in the same region of phase space. The approximate location of such region is clear and is labeled ``B'' in the bottom panel of Fig.\ \ref{fig.model_predictions1}(a), but corresponds to very different time instants in the time series (top panel).

The comparison between the reconstructed phase space (Fig. \ref{fig.old_expdata}(c)) and the model phase space (Fig. \ref{fig.model_simuls}(b)) is to be understood as a qualitative similarity in the geometrical arrangement of the trajectories: both have trajectories with asymmetric overshoot and both have trajectories that share the same region in phase space despite coming from opposite perturbations. Our model allows us to make the following prediction: a perturbation to the variable in the region labeled ``B'' in phase space (see Fig.\ \ref{fig.model_predictions1}(a)) should show the same post-perturbation time evolution no matter what original trajectory belongs to.

For the very same reasons exposed in Section \ref{sec.predicted}, in the history of paced finger tapping experiments it has been intrinsically difficult to perturb the value of the stimulus period $T_n$ without perturbing the variable $e_n$, and conversely to perturb the variable without perturbing the stimulus period. In a recent work, however, we showed the experimental feasibility of such manipulations \cite{Lopez2019}. We then propose to use these novel perturbations to study the response of the model in front of two consecutive perturbations:
\begin{enumerate}
\item first a traditional step change perturbation (in two conditions, positive and negative);
\item second, and while the resynchronization from the first perturbation takes place, a perturbation to the variable without changing the stimulus period.
\end{enumerate}

Our prediction is that the time evolution following the second perturbation will be the same for both conditions (positive or negative first perturbation), provided the second perturbation is performed when the system is approximately in the same region of phase space. The rationale behind this proposal is that, if we on the contrary performed two consecutive traditional step change perturbations, we would not be able to resolve the following confounder: in case an overshoot appears in both conditions after the second perturbation it may be due either because our hypothesis is valid or because it is the known response to the (second) step change perturbation.

Numerical simulations supporting our prediction are in Fig.\ \ref{fig.model_predictions1}, where we show the results of (a) a single step change (negative control), and the definition of perturbation points; (b) the proposed manipulation; (c) two consecutive perturbations as in (b) but performed in different points of phase space (positive control). In the consecutive perturbations (comparison (b)-(c)) the time evolution after the second perturbation is either very similar between conditions (b) or different (c).

\subsubsection{Perturbations to the variable only}

\begin{figure}[ht]
    \centering
  \includegraphics[width=0.5\textwidth]{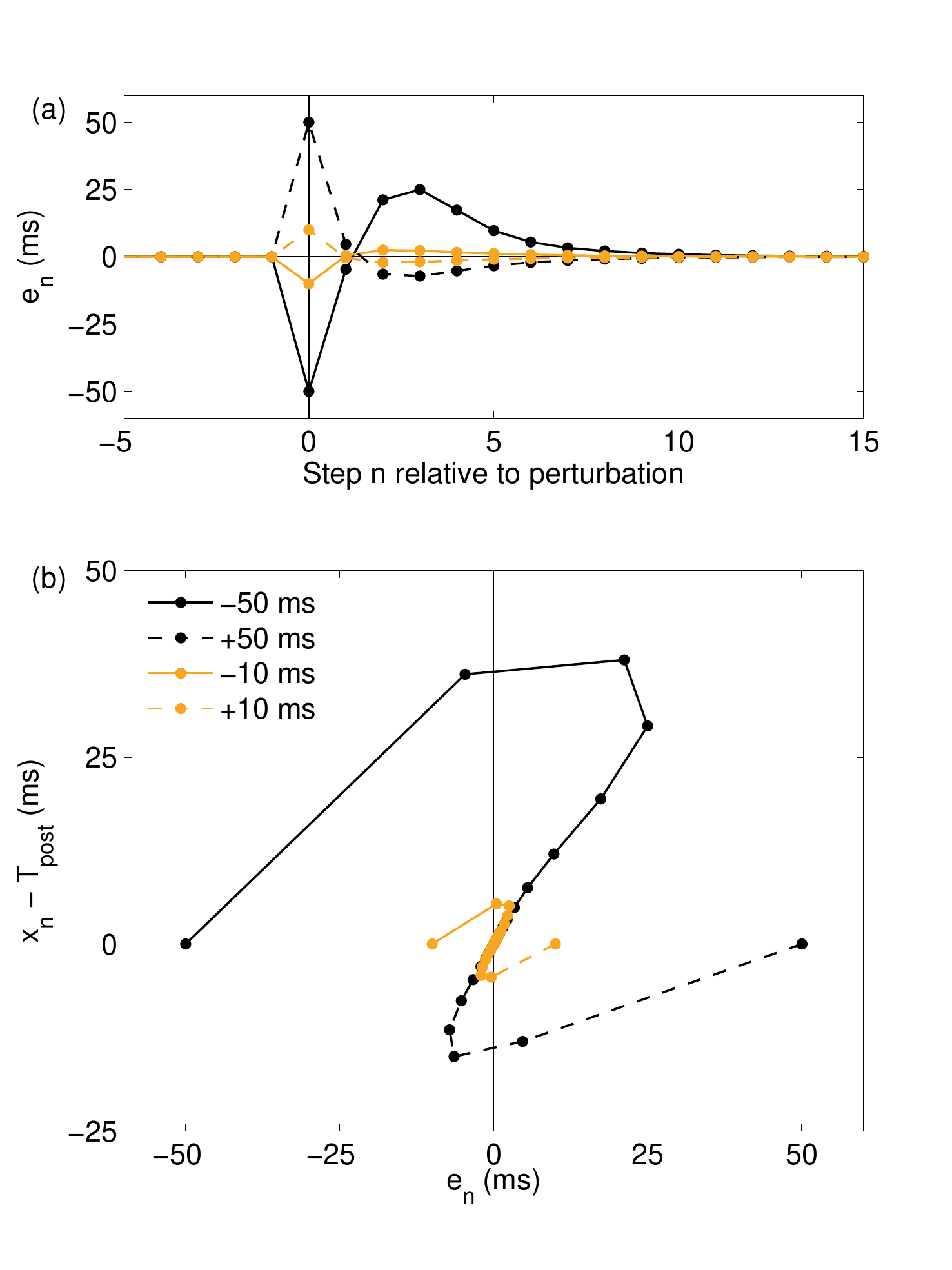}
    \caption{Model predictions (II).  Response to perturbations to the variable only (large: $\Delta T = \pm50$ ms; small: $\Delta T = \pm10$ ms). The response to large perturbations is asymmetric (i.e., larger overshoot in the negative perturbations); the response to small perturbations is mostly symmetric. (a) Asynchrony time series. (b) Trajectories in phase space.}
    \label{fig.model_predictions2}
\end{figure}

Our second prediction is that the response to large enough symmetric perturbations to the variable might be asymmetric. This can be seen in the phase space of Fig.\ \ref{fig.model_predictions2}, after noting that large negative perturbations (i.e., jump to the left) have a larger overshoot than large positive perturbations (i.e., jump to the right). This would prove that a step change perturbation is not needed to display nonlinear behavior---perturbations to the variable without changing the stimulus period might elicit it too. On the other hand, if the perturbation size is small enough then the response to symmetric perturbations might be symmetric (see Fig.\ \ref{fig.model_predictions2}).

\subsubsection{Large perturbations}
\label{sec.large}

We note the saddle node appearing in the upper right quadrant of the phase space, Fig.\ \ref{fig.model_simuls}(b). Its stable manifold is a separatrix between bound trajectories (i.e., converging to the origin) and unbound trajectories (i.e., diverging). The existence of the saddle node was not explicitly considered in the modeling effort; a post-hoc interesting interpretation is that it separates successful from unsuccessful resynchronization responses (bound and unbound trajectories, respectively). The unbound trajectories develop only when the perturbation is large enough, meaning the subject is not able to successfully resynchronize after large perturbations. Unbound trajectories display asynchronies with eventually diverging values; within the scope of our model, a very large asynchrony means that the response is to be asociated to the following (or previous) stimulus. Our model does not take into account this reassociation process and thus the validity of our description ends when the asynchrony crosses a threshold value, for instance half period.

Lastly, the predictions described above are valid if the perturbations are performed on time series displaying asymmetric overshoot, otherwise predictions and conclusions would be flawed.

\section{Conclusions}

We showed that the asynchrony $e_n$, the most important variable in the literature of paced finger-tapping experiments and theoretical models \cite{Chen1997,Repp2013}, is actually an ill-defined variable for a map or difference equation model when perturbations to the stimulus sequence are present. We proposed a distinction between predicted and actually observed asynchrony and developed the first mathematical model to solve the inherent ill-definition. This is also the first mathematical model in sensorimotor synchronization that takes into account the response to a temporal perturbation as a built-in feature. This is an important issue when considering interpersonal or group synchronization and leader-follower relationships (like in choirs and orchestras), as any naturally occurring variability in the timed actions of any participant or tempo change by the leader will act as a perturbation to the rest. Our own previous attempt \cite{Bavassi2013}, though successful at unifying, fell short of completely including the perturbation in the model dynamics.

Our model is able to fit the step change perturbation data remarkably well (Fig.\ \ref{fig.model_simuls}(a)), reproducing the response time series very accurately at all steps (including the perturbation step) with no modification of the variable ``by hand'' and with basically the same number of parameters than comparable models. It is only surprising that no published works so far in the paced finger tapping literature, up to our knowledge, deal with the issue of ill-definition of the main model variable when the sequence period is perturbed. On the other hand, models based on forced or coupled nonlinear oscillators \cite{Kelso1985,Large2002,Loehr2011,Jazayeri2019}, traditionally classified as belonging to a ``dynamical systems'' approach, are naturally well defined even in the presence of perturbations to the period.

Temporally displaced auditory feedback (either delayed or advanced) is also a usual way of probing the system \cite{AscherPrinz1997, Wing1977, PfordDalla2011, MatesAscher2000} and it does not suffer from the issue of ill-definition of the variable we addressed in this work. It remains to be shown, however, how it relates to changing the asynchrony in the models as it produces a modification in the asynchrony value but it also introduces a dissociation between auditory feedback and proprioceptive and tactile feedback \cite{PfordDalla2011}.

We also showed that nonlinear behavior (asymmetry of responses) might be observed when the variable only is perturbed, i.e.\ even in the absence of a perturbation to the stimulus sequence, if the perturbation is large enough. Experimental perturbations like the ones proposed in \cite{Lopez2019} but with larger magnitudes are needed. We acknowledge that similar results can be obtained by using a different set of nonlinear terms, and this calls for more experimental data showing any kind of bifurcation in the behavior so as to choose the correct set of parameters via normal form theory.

Our model assumes, as many others, that the origin of the asynchrony is not important for keeping average synchrony or for achieving resynchronization. Qualitative features of resynchronization after a perturbation are thus similar independently of whether the asynchrony was produced by a perturbation to the parameter or to the variable or both. This is a common theoretical assumption in the literature only recently supported by experimental results \cite{Lopez2019}.

Our theoretical results show that past (observed) and future (predicted) asynchronies play different roles in the model, and the remarkable fitting to the experimental data thus offers indirect evidence for a separate cerebral account of predicted versus actually observed asynchrony. Further experimental work is needed to decide whether this holds true.

\begin{acknowledgments}
This work was supported by Universidad Nacional de Quilmes (Argentina), CONICET (Argentina), and The Pew Charitable Trusts (USA).

Author contributions: CRG and RL analized data, wrote code and performed numerical simulations; RL conceptualized the work and developed the model; CRG, MLB and RL wrote the manuscript.
\end{acknowledgments}

\appendix

\section{Methods and Parameter distributions}\label{Methods and Parameter distributions}

\subsection{Estimation of return points}\label{Estimation of return points}

The return points of a map $e_{n+1} = f(e_n)$ are the values $e_n$ such that $e_{n+1} = e_n$ (the analogous concept in a continuous-time flow is the nullcline, that is the points in phase space where the rate of change associated to a given variable is zero). In the time series $e_n$ the return points appear as local maxima or minima.

\begin{figure*}[t]
    \centering
    \includegraphics[width=\textwidth]{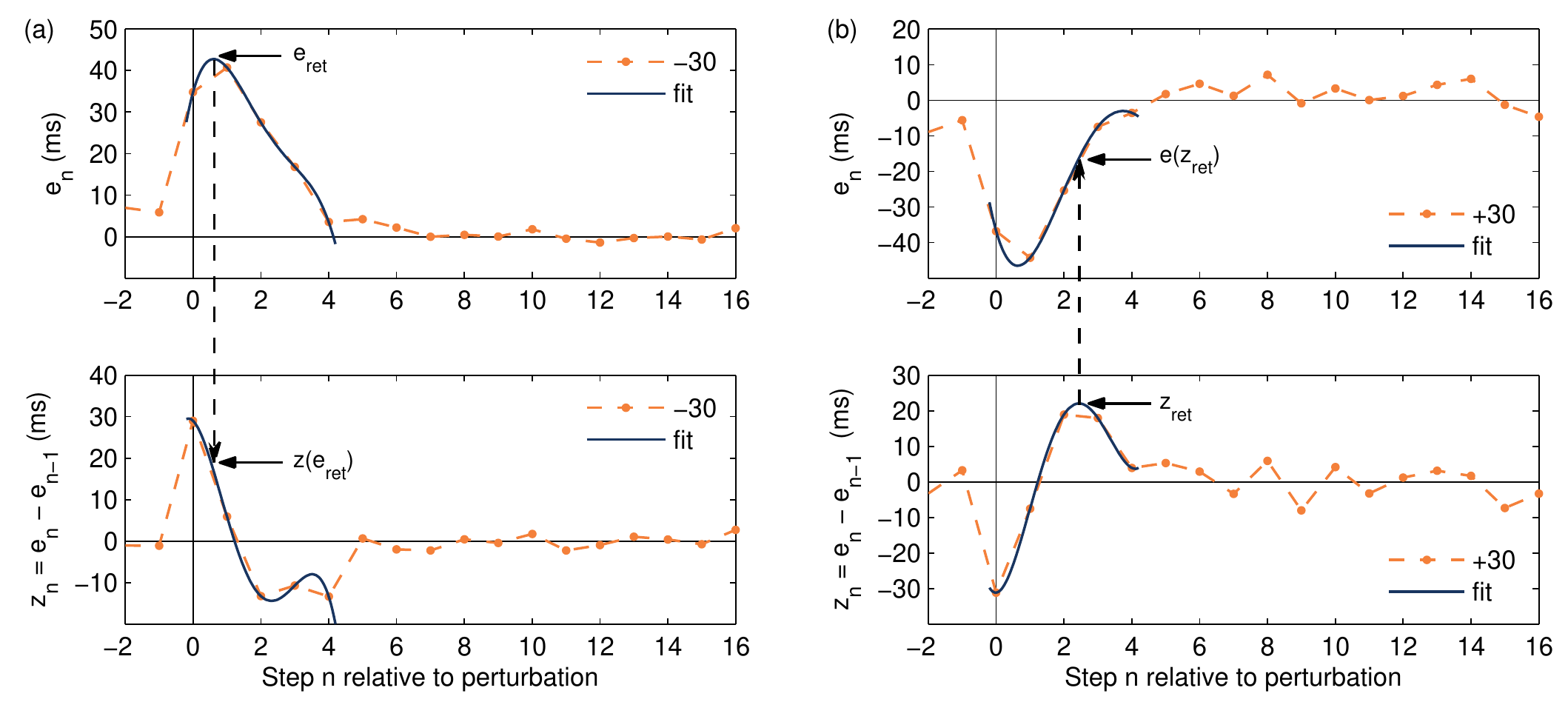}
    \caption{Estimation of return points. (a) Return points of $e_n$ (top panel; labeled ``horiz'' in Fig.\ \ref{fig.old_expdata}(c)) and corresponding $z_n$ value (bottom panel). (b) Return points of $z_n$ (bottom panel; labeled ``vert'' in Fig.\ \ref{fig.old_expdata}(c)) and corresponding $e_n$ value (top panel).}
    \label{fig.return}
\end{figure*}

In a 2D map:
\begin{equation}
\begin{aligned}
e_{n+1} & = f(e_n,x_n) \\
x_{n+1} & = g(e_n,x_n)
\end{aligned} \label{eq.return}
\end{equation}

\noindent the return points for the variable $e_n$ are the solutions of the implicit function $e_n = f(e_n,x_n)$ and for the variable $x_n$ are the solutions of the implicit function $x_n = g(e_n,x_n)$. Here it is clear that the return points give us information about the shape of $f$ and $g$.

In our case we want to find the return points in the chosen bidimensional embedding $(e_n ; z_n)$, where $z_n = e_n - e_{n-1}$. That is, on the one hand we want the points $(e_n ; z_n)$ such that $e_n$ is a local maximum or minimum (return points for the variable $e_n$), and on the other hand the points $(e_n ; z_n)$ such that $z_n$ is a local maximum or minimum (return points for the variable $z_n$).

We exemplify the procedure with the calculation for the variable $e_n$ (see Figure \ref{fig.return}):
\begin{enumerate}
    \item Define a 5-point time window from $n=0$ through $n=4$;
    \item Fit a 4th-order polynomial to the $e_n$ time series in such window;
    \item Find the local maximum or minimum $e_{\mbox{ret}}$ of the fitted function and the corresponding value $n_{\mbox{ret}}$;
    \item Interpolate the $z_n$ time series with a polynomial of the same order and compute the value $z(e_{\mbox{ret}})$;
    \item The return points are the set of values $(e_{\mbox{ret}};z(e_{\mbox{ret}}))$.
\end{enumerate}

Same procedure applies for the variable $z_n$ after switching $e_n \leftrightarrow z_n$.

\subsection{Genetic algorithm}\label{Genetic algorithm}

We fitted the model to the data by using a genetic algorithm in C with both custom-written code and the GAUL libraries (\url{http://gaul.sourceforge.net}).

The eight model parameters were arranged into a single chromosome with eight genes, and were initialized randomly from a uniform distribution in the ranges
$-1.0 < a,b,c,d < 1.0$, $-0.01 < \delta < 0.01$, and $-0.0001 < \alpha,\beta,\gamma < 0.0001$. The number of generations was 200, the population size 2000, the crossover rate $0.9$, and the mutation rate $0.1$.

The fitness function was defined as minus the square root of the average squared deviation between model series and experimental series:
\begin{equation}
F = -\sqrt{\frac{1}{2\times26} \sum_{j=1}^2 \sum_{n=-5}^{20}
\left( e_n^j - E_n^j\right)^2 + P}
\end{equation}

\noindent where $e_n^j$ is the experimental time series and $E_n^j$ is the model time series; $n = -5,\dots,20$ is the step number as in Fig.\ \ref{fig.model_simuls}(a), and $j=1,2$ represents the two conditions $\pm50$ ms used to fit the model. The fitness function $F$ decreases as the differences $e_n^j - E_n^j$ get larger in absolute value. Note that fitting and simulations in this work encompass all steps $n = -5,\dots,20$ in a sequence including the perturbation step $n=0$ as shown in Fig.\ \ref{fig.model_simuls}(a), making no distinction whatsoever among pre-perturbation, perturbation, and post-perturbation.

In order to prevent survival of unrealistic solutions (for instance damped oscillations or alternating series), penalties were included as a positive term $P$ inside the square root that depends on the linear coefficients only and takes a large value in any of the following cases:
\begin{enumerate}
\item the eigenvalues are complex (in order to avoid oscillatory approach to the equilibrium);
\item the eigenvalues are real but any of them is either greater than 1 or negative (in order to avoid solutions with unstable manifolds, and convergent solutions that alternate sides);
\end{enumerate}

\noindent otherwise $P=0$.

In order to prevent the selection of a surviving local optimum and to perform a statistical and dynamical study of the obtained solutions, the whole procedure described so far was repeated 200 times; the chosen solution was the one with the highest fitness of all.

To improve fitting, a post-perturbation constant baseline was added to the model variable $p_n$ with a fixed value
equal to the experimental post-perturbation baseline of the corresponding perturbation size.

\subsection{Fitted parameter distributions}\label{Fitted parameter distributions}

\begin{figure*}[ht]
    \centering
    \includegraphics[width=\textwidth]{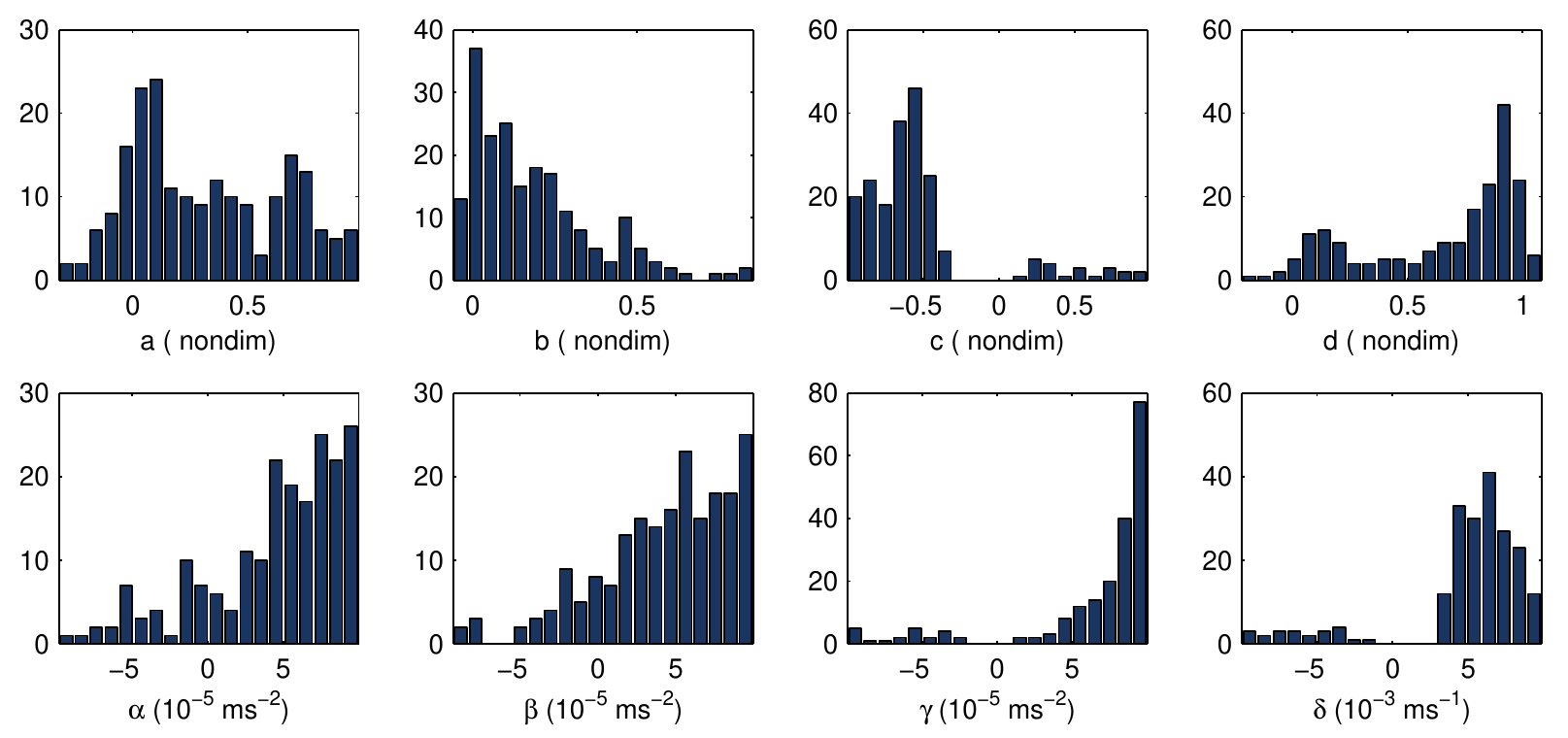}
    \caption{Distribution of parameter fitted values.}
    \label{fig.param_distribution}
\end{figure*}

\begin{figure}[h]
    \centering
    \includegraphics[width=0.5\linewidth]{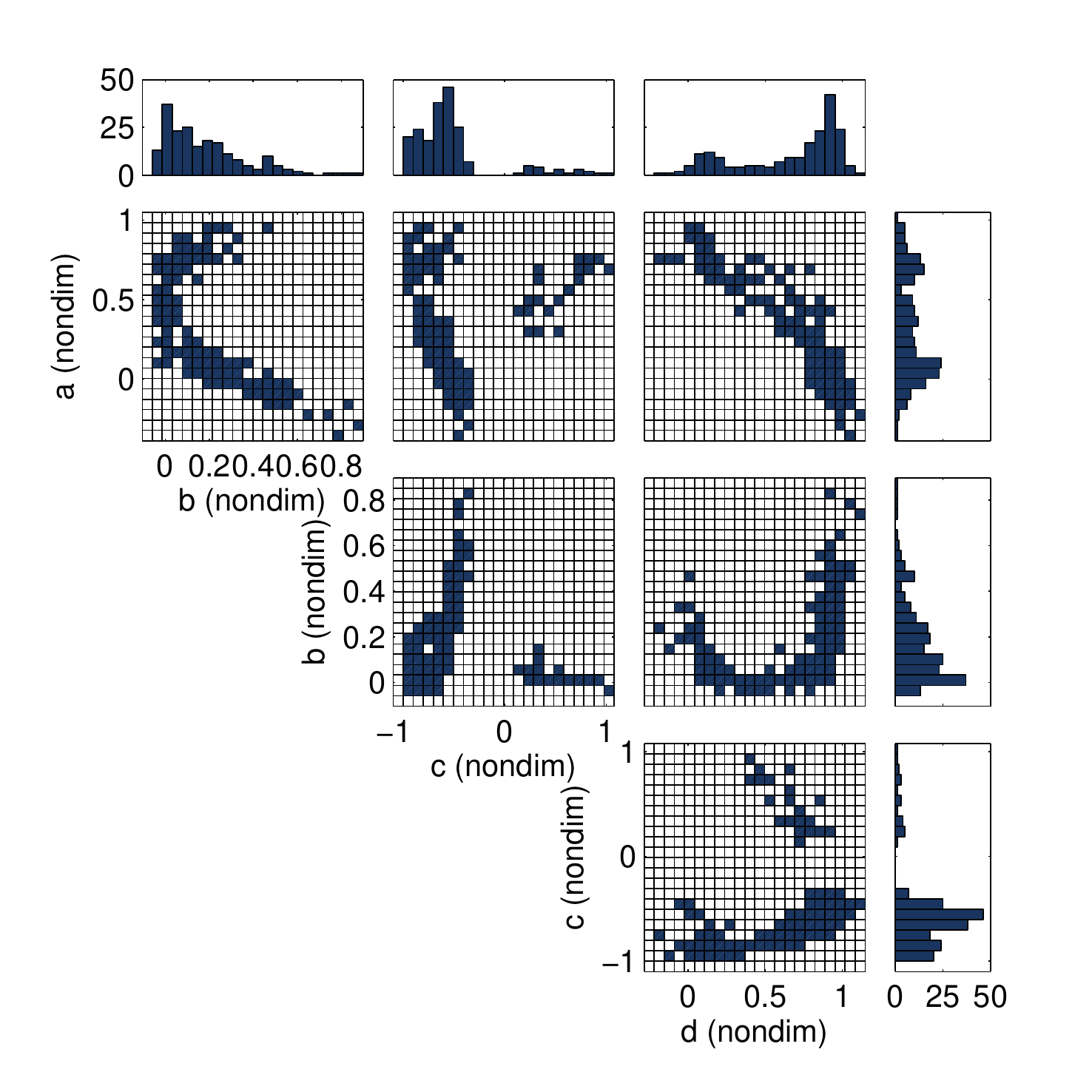}
    \caption{Joint distributions of linear parameter values.}
    \label{fig.param_distribution_linlin}
\end{figure}

\begin{figure}[ht]
    \centering
    \includegraphics[width=0.5\linewidth]{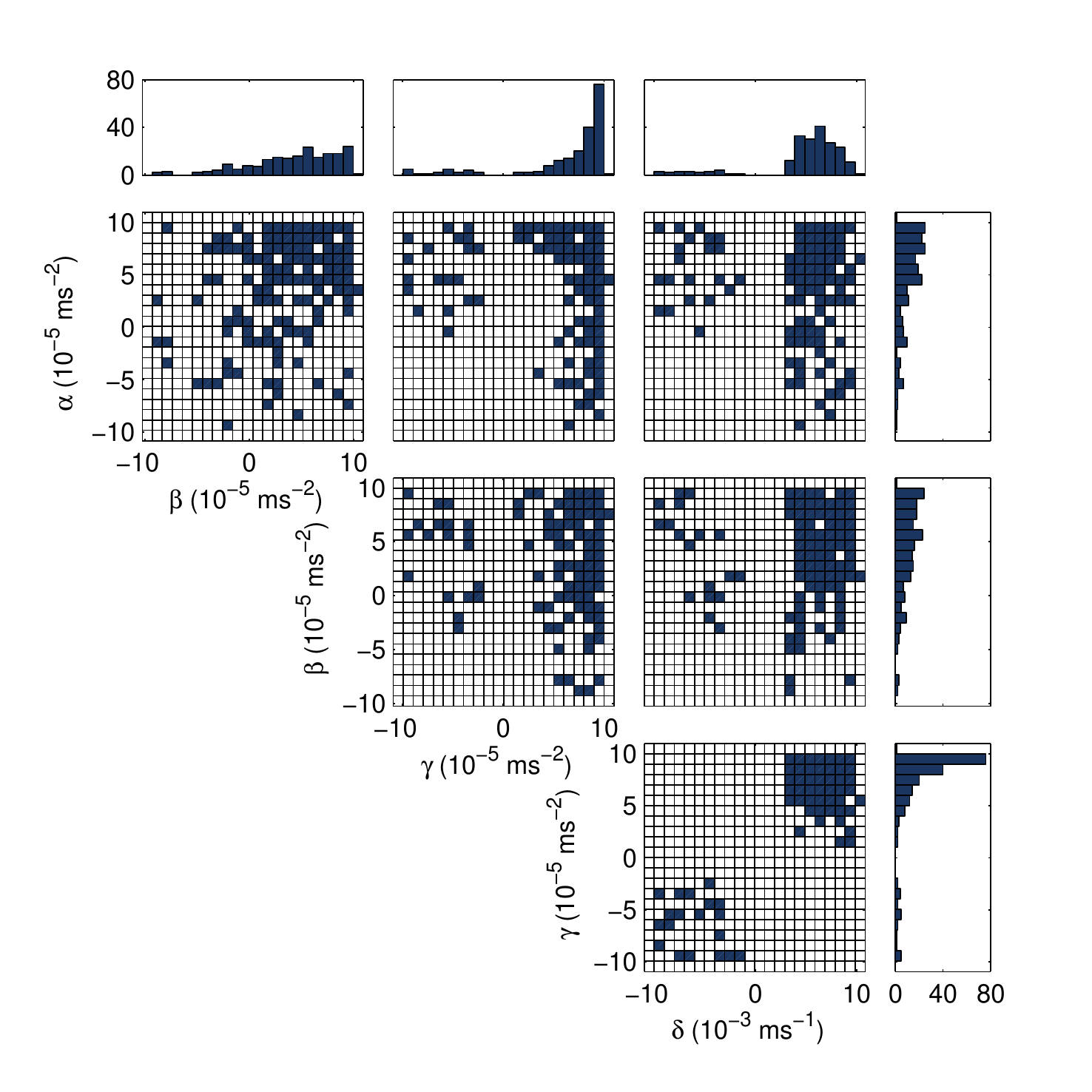}
    \caption{Joint distributions of nonlinear parameter values.}
    \label{fig.param_distribution_nolinnolin}
\end{figure}

\begin{figure*}[ht]
    \centering
    \includegraphics[width=0.75\textwidth]{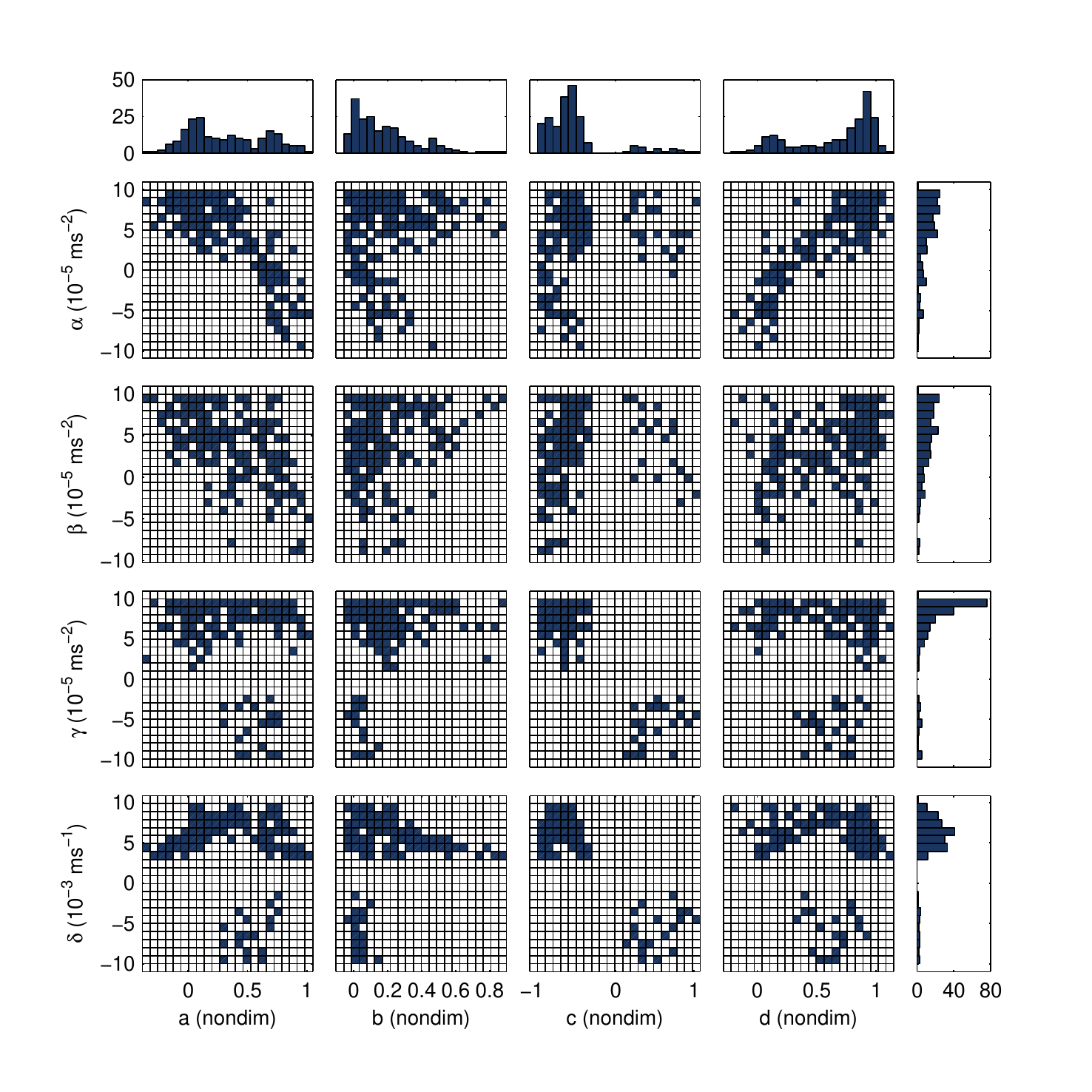}
    \caption{Joint distributions of linear vs nonlinear parameter values.}
    \label{fig.param_distribution_linnolin}
\end{figure*}

See Figures \ref{fig.param_distribution}, \ref{fig.param_distribution_linlin}, \ref{fig.param_distribution_nolinnolin}, and \ref{fig.param_distribution_linnolin}.

\subsection{Parameter values of Type-II and Type-III phase spaces}\label{Parameter values of Type-II and Type-III phase spaces}

\begin{table}[ht]
\centering
\begin{tabular}{rclp{0.25cm}|p{0.25cm}rcl}
$a$ & = & $0.943$ & & & $\alpha$ & = & $-1.15 \times 10^{-5}$ \\
$b$ & = & $0.176$ & & & $\beta$ & = & $-9.29 \times 10^{-5}$ \\
$c$ & = & $-0.842$ & & & $\gamma$ & = & $7.54 \times 10^{-5}$ \\
$d$ & = & $0.0690$ & & & $\delta$ & = & $3.50 \times 10^{-3}$
\end{tabular}
\caption{Fitted parameter values for a representative solution of Type-II phase space (Fig. \ref{fig.types}(b)). Units as in Table \ref{tab.param_values}.
\label{tab.param_values2}}
\end{table}

\begin{table}[ht]
\centering
\begin{tabular}{rclp{0.25cm}|p{0.25cm}rcl}
$a$ & = & $0.751$ & & & $\alpha$ & = & $4.34 \times 10^{-5}$ \\
$b$ & = & $0.0167$ & & & $\beta$ & = & $1.72 \times 10^{-6}$ \\
$c$ & = & $0.975$ & & & $\gamma$ & = & $-5.02 \times 10^{-5}$ \\
$d$ & = & $0.371$ & & & $\delta$ & = & $-3.80 \times 10^{-3}$
\end{tabular}
\caption{Fitted parameter values for a representative solution of Type-III phase space (Fig. \ref{fig.types}(c)). Units as in Table \ref{tab.param_values}.}
\label{tab.param_values3}
\end{table}

\subsection{Data and code}

See Supplemental Material \cite{SuppMat} at [URL will be inserted by publisher] or at the Sensorimotor Dynamics Lab's webpage: \url{www.ldsm.web.unq.edu.ar/perturbations2019} for C and MATLAB code to reproduce all figures and data in this work.

We use the morgenstemning colormap \cite{GeissbuehlerLasser2013} for color blind-friendly and grayscale-friendly plots.

\pagebreak

%\bibliographystyle{apsrev4-2}
%\bibliography{predicted_bib}

%apsrev4-2.bst 2019-01-14 (MD) hand-edited version of apsrev4-1.bst
%Control: key (0)
%Control: author (72) initials jnrlst
%Control: editor formatted (1) identically to author
%Control: production of article title (-1) disabled
%Control: page (0) single
%Control: year (1) truncated
%Control: production of eprint (0) enabled
%

\end{document}